\documentclass[10pt,letterpaper,twocolumn]{article} 
\usepackage{ol2}
\usepackage[draft]{hyperref} 
\usepackage{srcltx,flushend}

\begin{document}

\twocolumn[

\title{Experimental  study of $z$ resolution in Acousto-Optical Coherence Tomography
using random phase jumps on ultrasound and light}

\author{Max Lesaffre$^1$, Salma Farahi$^2$,  François Ramaz$^1$ and Michel Gross$^{3}$ }

\address{ $^1$Institut Langevin, ESPCI ParisTech, CNRS UMR 7587, 1, rue Jussieu F-75238 Paris  Cedex 05.}
\address{ $^2$School of Engineering, Ecole Polytechnique Federale de Lausanne (EPFL), Station 17, CH-1015 Lausanne, Switzerland .}
\address{  $^3$Laboratoire Charles Coulomb- UMR 5221 CNRS-UM2 Universit\'{e} Montpellier II place Eug\`{e}ne
Bataillon 34095 Montpellier}

\date{\today}
%

\begin{abstract}
Acousto-Optical Coherence Tomography (AOCT)  is  a variant of Acousto Optic Imaging (also called Ultrasound modulated Optical Tomography)  that makes possible to get resolution along the ultrasound propagation axis $z$. We present here new AOCT experimental results, and we study  how the $z$ resolution depends on
time step between phase jumps $T_\phi$,
or on the correlation  length  $\Delta z$. By working at low resolution,  we  perform a quantitative comparison of the $z$ measurements with the theoretical Point Spread Function (PSF).  We present  also images  recorded with different $z$ resolution, and we qualitatively show how the image quality varies with $T_\phi$, or $\Delta z$.
\end{abstract}


\maketitle

]

\section{Introduction}
%

\textbf{Citation:}

Max Lesaffre, Salma Farahi,  François Ramaz and Michel Gross
"Experimental  study of $z$ resolution in Acousto-Optical Coherence Tomography
using random phase jumps on ultrasound and light"
Applied Optics: \textbf{52},5, pp. 949 –- 957  (2013)

\bigskip

Acousto-optic imaging (AOI) \cite{leutz1995ultrasonic,wang1995continuous,kempe1997acousto,leveque1999utp,yao2000theoretical}  combines light and ultrasounds to measure local
optical properties through thick and highly scattering media.

The principle of this imaging method is the following. Ultrasound (US) is applied in the region of interest, within the
thick scattering sample, making  the scatterers
vibrate. A CW laser  illuminates the
sample. The vibration of the scatterers at the acoustic US
frequency $\omega_{US}$ (2MHz typically) modulates the phase of
the photons that are scattered by the sample. 
%
%
The light exiting the sample contains thus different
frequency components. The main component
(carrier) is centered at the laser illumination frequency $\omega_L$.
It is related to the scattered photons, that do not interact
with the US.
 The sideband components at frequencies
  $\omega_{\pm 1}$  are shifted from the US 
frequency so that $\omega_{\pm 1}=\omega_L \pm \omega_{US}$. The sideband photons, which result from the interaction between light and US, are called ''tagged photons'', their  weight  depends on the optical absorption in the region of interest, where the US beam is focused. Acousto-optic imaging detects selectively the tagged photons.
%
%
The signal is recorded while scanning the US within the volume of the sample, and thus one gets informations of the local optical absorption and/or scattering in 2D or 3D.
%


The first experiments were performed with  fast  single detectors  to record the  modulation of the optical signal at the US frequency
\cite{wang1995continuous,leutz1995ultrasonic,wang1997ultrasound,kempe1997acousto,yao2000theoretical}
, but, since the phase of the modulation is different for each grain of speckle, the
detector can only process one grain of speckle. In order to increase the optical etendue of detection, Leveque et al. \cite{leveque1999utp} have developed a camera coherent detection
technique that processes many speckles in parallel. This technique has been pulled to the
photon shot noise limit by Gross et al. \cite{gross2003snd} using a holographic
heterodyne technique \cite{leclerc2000nhh} able to detect photons with optimal
sensitivity \cite{gross2007dhu,verpillat2010digital}. But, because of the tissue inner motions,
the frequency  bandwidth of the photons that travel through living tissue is enlarged (1.5 kHz typically  through 4 cm of living breast tissue \cite{gross2005hdm}), while the bandwidth of the camera coherent detection ($1/T$) is limited by the exposure time $T$.
In order to increase this  bandwidth, one has to decrease $T$, and thus the detected signal remains almost the same. This means that the detected signal is essentially independent of camera exposure time $T$, if this time $T$ is  longer than correlation  time of the sample speckle. This point has been experimentally demonstrated with turbid liquid \cite{atlan2007spatiotemporal}.
An alternative to match the detection and the signal bandwidth while keeping a large optical etendue is to use a detection involving
%
 a detection bandwidth comparable  with the  signal bandwidth while keeping a large optical \\
etendue, is to use a detection schemes involving
photorefractive (PR) crystals  \cite{murray2004detection,sui2005imaging,ramaz2004photorefractive,lesaffre2007smp,xu2007photorefractive}.

Since the US attenuation is low in tissues, tagged photons are generated along the US
propagation $z$ axis at nearly constant rate. This means that in a continuous regime
of the US, the AO techniques give nearly no information on the location of the embedded
objects along the $z$ axis. In order to get $z$ information, acoustic
pulses can be used whether with  single detectors
\cite{lev2003pulsed,lev2005ultrasound}, cameras \cite{atlan2005pulsed} and  PR crystals
\cite{murray2004detection,sui2005imaging,ramaz2004photorefractive,lesaffre2007smp,farahi2010photorefractive,rousseau2008umo}. Nevertheless, reaching a millimetric
resolution with US pulses requires a typical duty cycle of $1\%$, corresponding to the ratio of the
exploration length within the sample ($\sim 10$ cm) and the desired resolution ($\sim 1$
mm). This is problematic regarding the very small quantity of light that emerges from a
clinical sample, since a weak duty cycle yields a low signal and a poor SNR.

%
%
%
%
%
%
%
%
%
%

In a recent publication, Lesaffre et
al. \cite{lesaffre2009acousto} overcome the duty cycle problem, and  get  $z$ resolution with a variant of the acousto optic technique, called Acousto Optical Coherence Tomography (AOCT). In this technique,  the same random phase modulation is applied on both light and US, but the  light modulation is applied at retarded time $t+z/c_{US}$, where $z$ is the coordinate of the selected zone, and $c_{US}$  the US velocity. The Lesaffre et
al.  experiment has been made with PR detection of the tagged photon signal, and, in that case, the exact shape of the $z$ instrumental response  has been calculated \cite{lesaffre2011theoretical}.

In this paper, we will present new experimental results in order to verify this theory by studying  how the $z$ resolution depends on the characteristic time of the random phase jumps, $T_\phi$. We will recall briefly the theoretical results of reference  \cite{lesaffre2011theoretical}, and introduce the AOCT resolution $\Delta z$. By working at low resolution,  we will perform a quantitative comparison of the measured  $z$ shape, with the  theoretical one. We will then present  images  recorded with different resolution $\Delta z$, and qualitatively show how the images quality varies with $\Delta z$.

\section{The experimental setup}

\begin{figure}[]
\begin{centering}
\includegraphics[width=8.5cm]{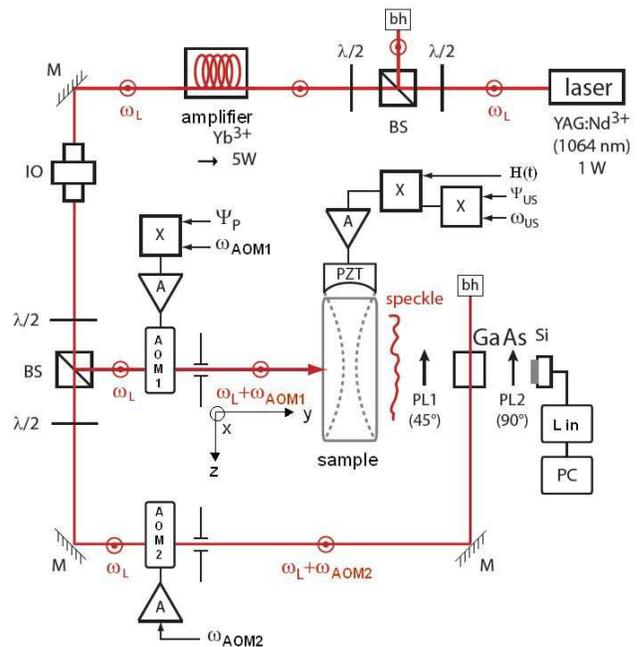}
\caption{AOCT experimental setup. Laser: Nd: YAG laser (1 W @ 1064nm); amplifier: optical amplifier 5W-doped Yb; IO: optical isolator Faraday; $\lambda/2$: half-wave plate; AOM1, 2: acousto optic modulators; M: mirror; BS: polarizing beam splitter; X : Radio Frequency (RF) $\sim$ 80 MHz Double
Balanced Mixer; A: RF amplifier; GaAs: photorefractive GaAs crystal; PL1,2: linear polarizers; Si: Silicon photodiode (0.3 cm$^{2}$); L in: Lock-In amplifier; PC: personal computer; PZT: Piezoelectric acoustic transductor ($\sim$ 2 MHz).  H(t): $0,\pi$ random phase modulation  at frequency $ \omega_{mod}$ = 3kHz, and with duty cycle $r = 0.24$; $ \psi_{US}$, $\psi_{P}$: random sequence of phase.}
\label{fig_setup}
\end{centering}
\end{figure}

The experimental setup is presented on Fig.\ref{fig_setup}. It is close to the one described elsewhere \cite{lesaffre2009acousto}.
An  acoustic transducer (Panametrics A395S-SU, focal length of 75 mm) generates a US beam, which is focused on the sample (frequency $\omega_{US}=2.3$ MHz, with a maximum US pressure $P_{US} \simeq 1.6 $ MPa at the focal point).
The master laser  (wavelength $\lambda$ = 1064nm, 5 W after amplification) is spatially single mode (transverse and longitudinal) and vertically polarized. It is splitted in two beams, whose amplitude, phase and frequency are controlled by two acousto optic modulators (AOM1 and AOM2) driven at $\omega_{AOM1}\simeq \omega_{AOM2}=80$ MHz. The first beam
illuminates the sample (area $s \sim 1$ cm$^2$). Note that the output signal beam is totally depolarized  because of the multiple diffusion events in the scattering sample. The second \\
  (reference) pumps  the photorefractive (PR) crystal. The relative power
of the reference versus illumination beam is adjusted by a half wave plate ($\lambda/2$) followed
by a polarizing beam splitter (BS). The PR crystal is a bulk GaAs crystal ($1.4 \times
1.6 \times 1.4$ cm$^3$ \cite{lesaffre2009acousto}), oriented into an anisotropic diffraction configuration
\cite{chang1988cpp,rousseau2008umo}. The signal beam, which is depolarized, and the reference beam, which is vertically polarized, enter onto orthogonal faces, respectively (11$\sqrt{2}$) and
(11$\overline{\sqrt{2}}$), in order to have a grating vector along $\langle 001 \rangle$.
In such a configuration, the effective electro optic coefficient is $r_{eff} = r_{41}$. A
polarizer oriented at $\pi/4$ relative vertical is positioned after the sample, while a
horizontally oriented  polarizer is positioned after the crystal. This configuration
minimizes  scattering of the intense reference beam  by the PR crystal, which
corresponds to the main parasitic light contribution. For a typical reference beam of
power $P_R=650$ mW over an area $S_R=0.32$ cm$^2$, the photorefractive response time is
$\tau_{PR} = 3.4$ ms.

\begin{figure}
  \begin{center}
  \includegraphics[width=8.5cm]{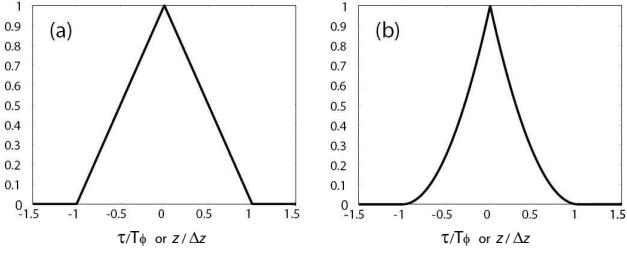}\\
  \caption{Plot of correlation function  $\underline{g}_{1}(\tau)$ (a), and  its square  $\underline{g}_{1}^2(\tau)$ (b).
The horizontal axis units is either $\tau/T_{\Phi}$ (for time correlation), or $z/\Delta z$ with $\Delta z=  c_{US}~T_{\Phi}$
(for $z$ resolution).}\label{Fig_correlation}
 \end{center}
\end{figure}

In order to perform  the $z$ axis selection of the tagged  photons, the phases of
illumination ($\psi_P (t)$) and  US ($\psi_{US} (t)$) beams  are pseudo-randomized.
Every time interval $T_{\Phi}$, we  apply pseudo random phase  jumps of $0$ or $\pi$ on both illumination and US (i.e. we multiply the corresponding sinusoidal signal by either +1 or -1 every $T_{\Phi}$).
By using two DBM (Double
Balanced Mixers ZAD-1H: Mini-Circuits Lab. Inc.), the sine wave signals of frequency
$\omega_{AOM1}$  and $\omega_{US}$ that drive the acousto optic modulator AOM1 and the
PZT transductor  are mixed with two correlated rectangular pseudo random phase signals $e^{j
\psi_P (t)} = \pm 1$ and $e^{j \psi_{US}(t)} = \pm 1$. These pseudo random  signals are generated
by two 80 MHz arbitrary waveform generators (Agilent 33250A) with a common 10 MHz
reference clock.
Because of the finite memory ($N$ words) of the generators, the phase
signals, which make random jumps at time intervals of $T_{\Phi}$, are periodic (pseudo
random) with periodicity $T=N T_{\Phi}$, where $N=2^{14}$ typically. In order to characterize the random phase sequence, one can introduce the phase correlation function $\underline{g}_{1}$ \cite{lesaffre2011theoretical}:
\begin{eqnarray}
\underline{g}_{1}(\tau) &=& \langle e^{j \psi_P (t)} e^{-j \psi_P (t-\tau)}   \rangle \\
\nonumber &=&  \langle e^{j \psi_{US} (t)} e^{-j \psi_{US} (t-\tau)}   \rangle
\end{eqnarray}
Since $N \gg 1$, the  correlation function $\underline{g}_{1}(\tau)$ has a triangular shape whose Full Width at Half Maximum (FWHM) is  $T_{\Phi}$ in time units, and $\Delta z= c_{US} T_{\Phi}$ in length units (see Fig. \ref{Fig_correlation} (a) ).

In order to choose the $z$ selected zone in the sample, the phase  $\psi_P $ is delayed with respect to
$\psi_{US}$ in a way that
\begin{eqnarray}\label{eq phi p phi us}
\psi_P (t) = \psi_{US} (t -\tau)
 \end{eqnarray}
with $ \tau = z/c_{US}$.\\

The signal detection procedure  is the same as  in reference
\cite{lesaffre2009acousto}. In order to get a time varying signal on the photodiode PD,
we apply on the US beam  an extra $H(t)=\pm 1$ rectangular modulation signal  at frequency $\omega_{mod}/2\pi = 1/T_{mod}=3$ kHz, with duty cycle $r$.
This frequency has been chosen to satisfy the condition of validity of theory  (see Fig. 2 of reference \cite{lesaffre2011theoretical}),
\begin{eqnarray}\label{eq cond_th}
T_\Phi \ll T_{mod} \ll \tau_{PR}.
\end{eqnarray}
Other modulation frequencies are  possible if    Eq. \ref{eq cond_th} condition  remains fulfilled. We have chosen to use here a $\pm 1$ rectangular modulation, because this modulation is (i) easy to make, (ii) yields a quite good Lock-in signal, and (iii) has been considered in theory \cite{gross2005theoretical,lesaffre2009acousto}.

Within the interval $t=0$ to $T_{mod}$ we have thus:
\begin{eqnarray}
 H(t)&=&+1 ~~~~ \textrm{for}~~~~ 0< t/T_{mod}< r  \\
 \nonumber     &=&-1 ~~~~\textrm{ for}~~~~ r< t/T_{mod}<1
\end{eqnarray}
The US beam is $H(t)=\pm 1$ modulated by using a second DBM in an AM modulator configuration. A US  signal at $\sim 2 $ MHz is feeded within the LO port of the DBM, while the $H(t)$ rectangular signal  is feeded within the IF port. The $H(t)$ signal is generated by a computer driven signal generator, which controls the frequency $\omega_{mod}/2\pi$ and the duty cycle $r$.  The $\pm 1$  modulated US output is then  obtained on the LO port.  Since the modulation is fast with respect to the recording of the tagged photon grating on the PR crystal
($\omega_{mod}\gg 1/\tau_{PR}$), the modulated component of the index grating that is
grooved in the PR crystal is proportional to $[(1-2r)r\textrm{sinc}(\pi r)]$ where
$r$ is the extra modulation duty cycle \cite{gross2005theoretical}. In order to maximize this
component in the Lock-in detection, the duty cycle   is fixed at $r=0.24$.

The  photodetector (PD) is a silicon photodiode (Hamamatsu S2386-8k, area $S_{PD}=0.3$
cm$^2$) connected to a home-made 10M$\Omega$ transimpedance amplifier.  The  modulated component of
the Photodiode signal at $\omega_{mod}$ is measured by a Lock-in amplifier (EG$\&$G Inc.
7210; integration time $\tau \simeq 100$ ms).
We have stacked all the (sample+crystal+photodetector) together in order to optimize the detection flux without using lenses.

%

\section{Quantitative study of the AOCT Point Spread Function (PSF) along $z$}\label{section_z_quant_study}

\begin{figure}[]
\begin{center}
    \includegraphics[width=8.5cm]{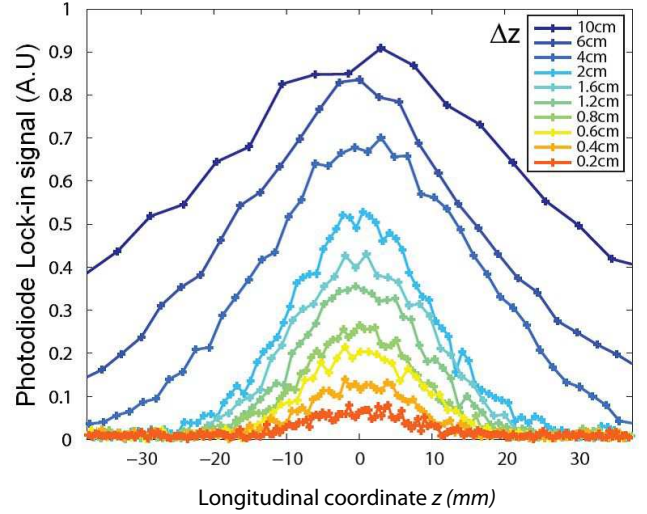}
\caption{Photodiode Lock-in acousto optic signal as a function of the longitudinal coordinate $z$ for
 for $ \Delta z =$ 10.0, 6.0, 4.0, 2.0, 1.6, 1.2, 0.8, 0.6, 0.4 and 0.2 cm.}
\label{fig_fig_curve_PD_signal_z}
\end{center}
\end{figure}

Lesaffre et al. \cite{lesaffre2011theoretical} have calculated the Point Spread Function (PSF) of the AOCT  signal, and show that $ \textrm{PSF} (z)$ is simply equal to $ \underline{g}_{1}^2$ expressed in length units. This function is plotted on Fig.\ref{Fig_correlation} (b), and corresponds to what expected in experiments. In order to measure  $ \textrm{PSF} (z)$, and to compare it with theory, we have plotted the photodiode Lock-In Signal $S_{PD}$ as a function of the delay $\tau$  for  phases  sequences with different correlation times $T_{\Phi}$. By this way, one can vary the location $z= c_{US} \tau $ of the selected zone, and the correlation length $\Delta z= c_{US} T_{\Phi}$. The experiment is made with  an \textit{Agar + Intralipid} 10\%  phantom sample without absorbing inclusion. The  reduced scattering coefficient is $\mu'_S=10$ cm$^{-1}$. The sample thickness  is $L_y= 3 $ cm along the illumination direction $y$. Other dimension are $L_x \times L_z= 6.5 \times 6.5$ cm$^2$.   The optical power is $P_S=1$ W/1 cm$^2$ on the sample, while the reference beam on the GaAs crystal is  $P_R=600$ mW/0.32 cm$^2$. Acoustic pressure at maximum is $P_{max}=1.6 $ MPa, and lock-in integration time is $\tau_c=100$ ms.

Although the experiments presented is this paper use the same setup, they have been done at different time, with slightly different laser power and optical adjustments. The parameters $P_S$, $P_R$,  $P_{max}$ and $\tau_c$ may thus change from one experiment to the next. Similarly, the sample parameters: thickness $L_y$, lateral dimension $L_x, L_z$ and scattering coefficient $\mu'_S$  may have changed, for good or bad reasons.

\begin{figure}[]
\begin{center}
    \includegraphics[width=4.2cm]{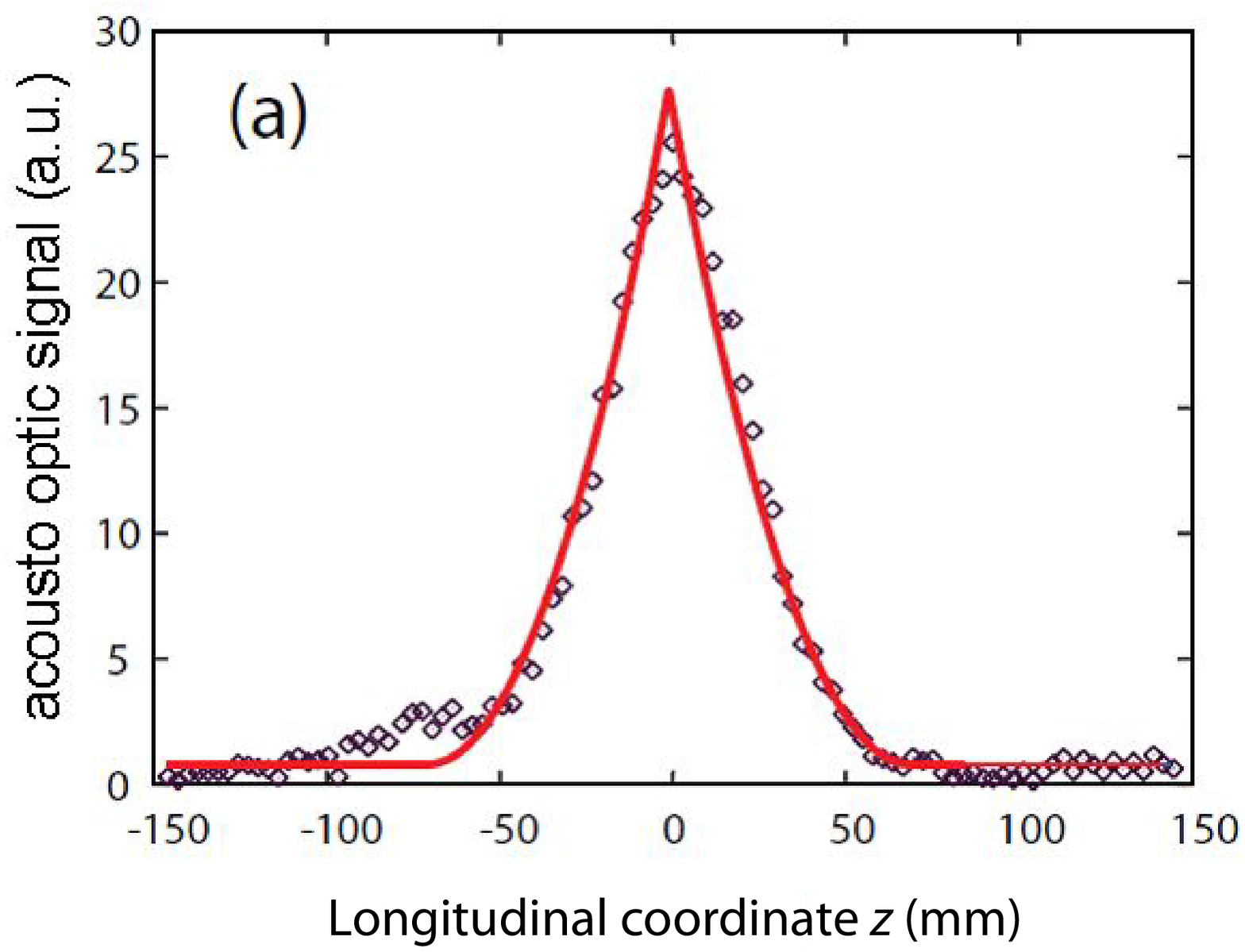}
   \includegraphics[width=4.2cm]{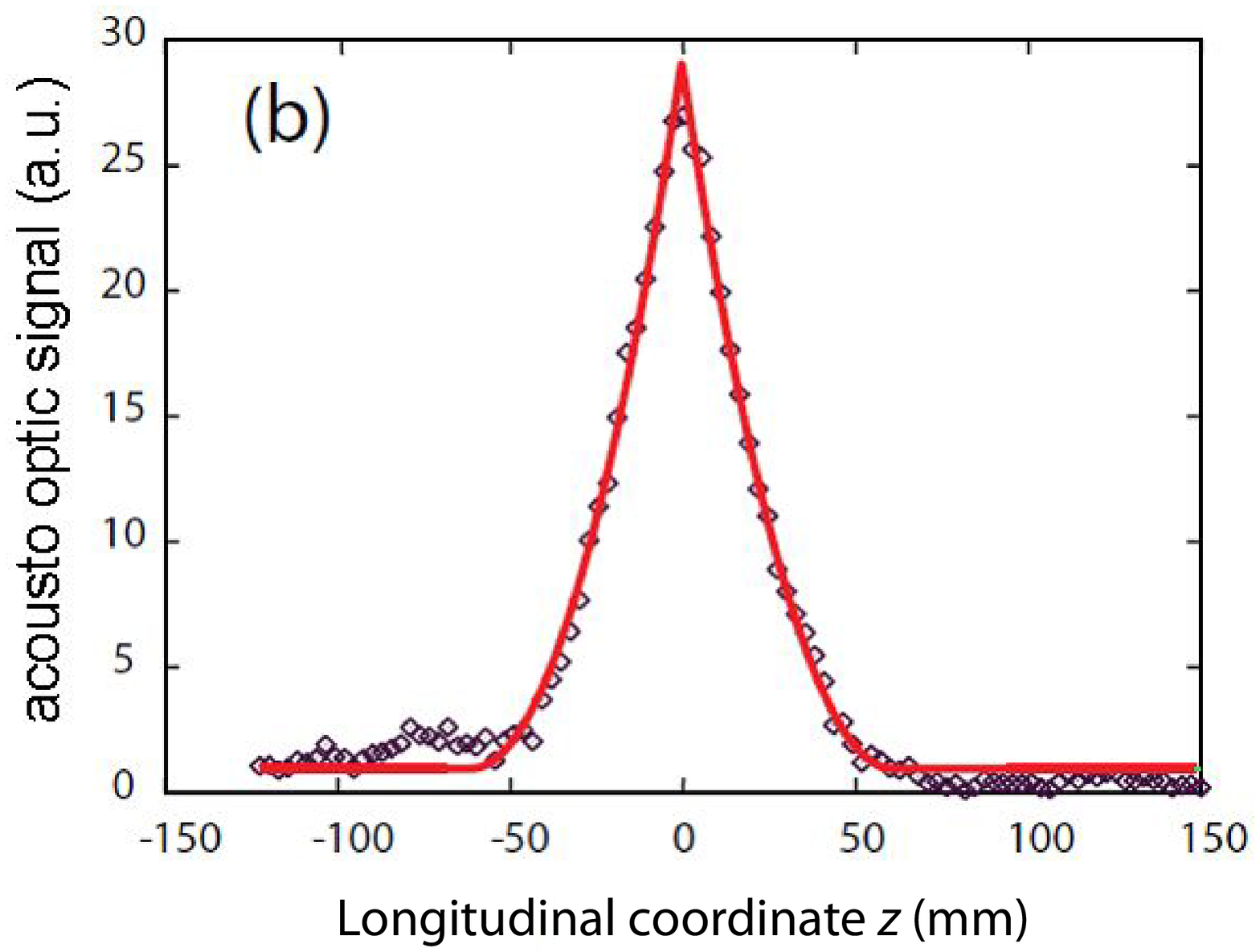}
\caption{Experimental profiles (black circles) made with $ \Delta z =$  10.0  (a), and   6.0  cm (b).   Theoretical shape $ \underline{ g1}^2 $ (red curves) made  by adjusting $ \Delta z $ in order to get the best  fit with experiment.  Fit yields   $\Delta z =10.27$ (a) and 6.09 cm (b).}
\label{fig_4.11}
\end{center}
\end{figure}

Figure \ref{fig_fig_curve_PD_signal_z} shows the Photodiode Lock-In signal  as a function of $z$  for $ \Delta z $ varying from  10 cm down to 
0.2 cm.  When $ \Delta z$ is small (0.6, 0.4 and 0.2 cm), the curve maximum is almost proportional to $ \Delta z$, while  the width remains nearly constant. The shape of the curves corresponds then to the shape, along $z$, of the  cloud of photons that travel through the sample and that reach the Photodiode. In the region  crossed by the US beam, the thickness of the cloud  along $x$ and $z$  is about 1.5 cm i.e. $ 0.5 \times $ the thickness of the sample.   When $ \Delta z$ is larger than the width of the cloud, i.e. for $\Delta z=$ 6 and 4 cm, the curve maximum saturates, and the width increases linearly with $ \Delta z$. In that last case,  one  can neglect the width of the cloud, and the $S_{PD}(z)$  corresponds  to $\textrm{PSF}(z)$.

To make a first quantitative test of theory \cite{lesaffre2011theoretical}, we have plotted both the measured acousto optic signal and the theoretical point spread function (PSF) ${g_1}^2$. We have considered situation where the PSF width  ($\Delta z=10 and 6 cm$)  is larger than the thickness of the cloud (about 1.5 cm). In such case, the experimental profile must be close to theoretical PSF ${g_1}^2$.  We has thus plotted ${g_1}^2$ (red curves of Fig. \ref{fig_4.11}) by adjusting $\Delta z$ in order to get the best fit with the experimental data . The fit yields $\Delta z= 10.27 and 6.09$ (for the ${g_1}^2$ curves) in excellent agreement with  $\Delta z=10 and 6 cm$ (experiments).

\begin{figure}[]
\begin{center}
    \includegraphics[width=5.5cm]{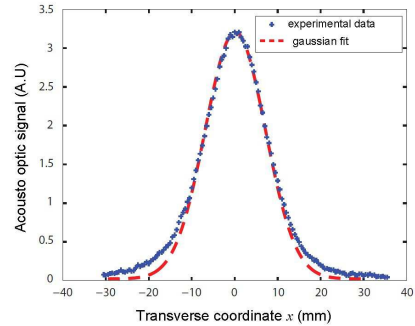}\\
\caption{Photodiode Lock-In  signal $S_{PD}$ as a function of the PZT location along the $x$ axis. Experimental data (cross), and  $\exp({-x^2/w_g^2})$ Fit with $w_g = 10.2$ mm (red dashed line).}
\label{fig_cond_exp_resolution_axiale_1D}
\end{center}
\end{figure}

In order to compare the $S_{PD}(z)$ experimental curves with theory for any $ \Delta z$, one has to calculate $ \underline{g_1}^2(z) \otimes C_z(z)$,  where $ \otimes$ is the convolution operator, and $C_z(z)$ the shape of the cloud of photons along $z$. Since the cloud  is symmetric with respect to the photon propagation direction $y$, the shapes  $C_x$ and $C_z$ along $x$ and $z$  are the  same.
We have measured  $C_x$ by moving  the PZT transducer along $x$ with a step motor, and recorded the photodiode  signal $S_{PD}(x)$ as a function of the PZT location $x$.   $S_{PD}(x)$ is plotted on  Fig.\ref{fig_cond_exp_resolution_axiale_1D}.

\begin{figure}[]
\begin{center}
    \includegraphics[width=8.5cm]{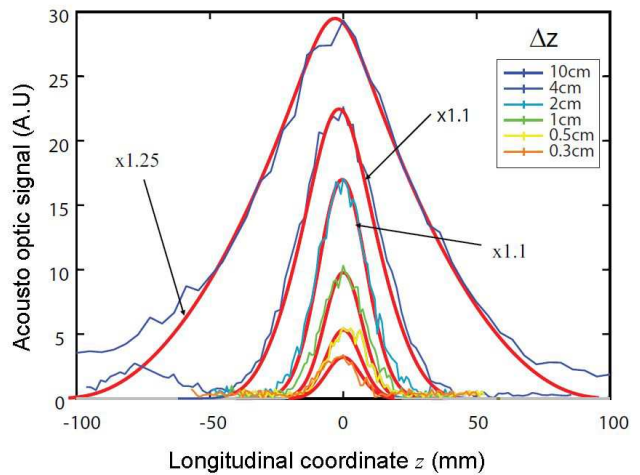}\\
\caption{Colored curves: experimental profiles $S_{PD}(z)$; Red curves: theoretical profiles $A \underline{g_1}^2(x) \otimes \exp(-x^2/w_g^2)$ where $ \otimes$ is the convolution operator. $\Delta z $ is 10, 4, 2, 1, 0.5 and 0.3 cm.}
\label{fig_4.12}
\end{center}
\end{figure}

We have  assumed that $C_z(x) = C_x(x) = S_{PD}(x)$, and  we approximate $C_x(x)$ by this gaussian fit $C_x(x) \simeq e^{-x^2/w_g^2}$ with $w_g = 10.2$ mm (see Fig.\ref{fig_cond_exp_resolution_axiale_1D}).
Replacing  $C_z(z)$ by the fit, we have calculated  the theoretical profiles $A \underline{g_1}^2(x) \otimes \exp(-x^2/w_g^2)$, where $A$ is an arbitrary gain parameter that should be the same for all value of $\Delta z$. To compare theory and experiment,  we have adjusted the gain parameter $A$ in order to get the best agreement. As mentioned on the curves of Fig.  \ref{fig_4.12}, we have slightly modified the gain parameter replacing $A$ by $1.1 \times A$ for $\Delta z = 4$   and  2 cm, and by $1.25 \times A$ for   $\Delta z = 10$ cm.  This small change of the  gain parameter is interpreted here as small drifts  (laser power,...) yielding small changes of the experimental signal and/or  detection efficiency during the several hours of data acquisition.
We have displayed   the theoretical and experimental profiles  on Fig.\ref{fig_4.12}. The agreement is excellent.

The $S_{PD}(z)$ curves displayed on Fig. \ref{fig_4.11} and Fig. \ref{fig_4.12} exhibit two regimes. The first regime is observed when $ \Delta z$ is smaller  than the width of the cloud of photons of Fig.\ref{fig_cond_exp_resolution_axiale_1D}
i.e. for  $\Delta z < 1.5$ cm.  The width of the curve  remains nearly constant, while the curve maximum is almost proportional to $ \Delta z$. This is expected since in that case the width of the cloud is larger than PSF (i.e. for $\Delta z > 1.5$ cm). The shape of the curves corresponds then to the shape of the  cloud  along $z$. On the other hand, the AOCT signal (i.e. the area  of the Fig. \ref{fig_fig_curve_PD_signal_z} curves) is ever proportional to the number of diffuser that contribute coherently to the AOCT signal (i.e. is proportional to $\Delta z$). As a consequence,  since the width remains almost constant (and is equal to the width of the cloud),  the maximum is proportional to $ \Delta z$.

The second  regime is observed when $ \Delta z$ is larger than the width of the cloud, i.e. for $\Delta z> 1.5$cm. The curve maximum saturates, and the width increases linearly with $ \Delta z$. This is expected since in that case the width of the cloud is smaller than PSF (i.e. for $\Delta z > 1.5$ cm). One  can then neglect the width of the cloud, and the Fig. \ref{fig_fig_curve_PD_signal_z} curves  corresponds  to the   width of $\textrm{PSF}(z)$, which increases like $ \Delta z$, and thus like the AOCT signal.  The maximum of the signal then saturates.

\section{AOCT images along the $x$ and $z$ directions.}

 AOCT has been used to obtain 2D images of different samples. The random phase jump time $T_{\Phi}$ is kept constant. The scan along $z$ is made by varying the AOCT delay $\tau$ with $z= \tau C_{US}$. The scan along $x$ is performed by moving the location $x$ of the PZT transducer with a motor. In order to reduce the noise, the 2D acquired data are low pass filtered in the Fourier space with a cutoff frequency $k_{c}$.

 \subsection{Image of the cloud of Photon: sample without absorbing inclusion.}

\begin{figure}[]
\begin{center}
    \includegraphics[width=6.0 cm]{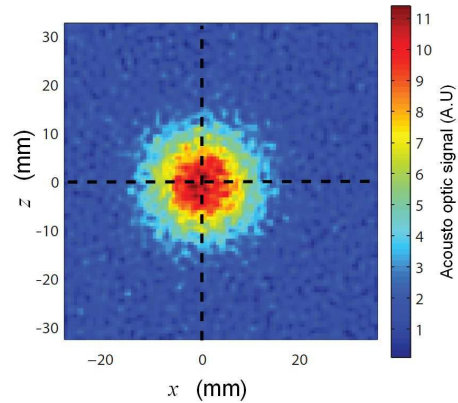}
\caption{AOCT  image ($x,z$) of a scattering sample without absorbing inclusion
($e= 3.2$ cm, $ \mu'_s =$ 6 cm$^{-1}$).}
\label{fig_image_banana}
\end{center}
\end{figure}

\begin{figure}[]
\begin{center}
    \includegraphics[width=8.5 cm]{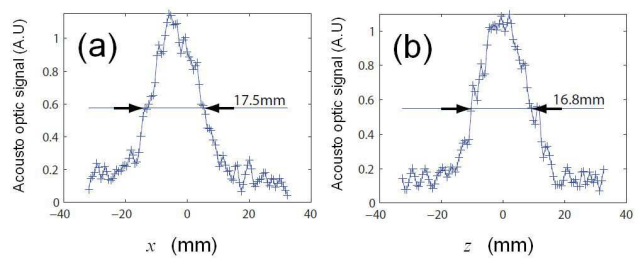}
\caption{Profiles of the AOCT photodiode lock-in signal  of the sample of Fig.\ref{fig_image_banana} along $x$ (a) and $z$ (b) directions.}
\label{fig_cuts_banana}
\end{center}
\end{figure}

To validate the analysis of the PSF made in section \ref{section_z_quant_study}, we have consider first a sample without inclusion in order to image the cloud of photons itself.

We have first imaged a sample without absorbing inclusion. The experiment is made with a $\mu'_s=6$ cm$^{-1}$ scattering sample, whose  thickness is $L_y=3.2$ cm, and whose  dimensions are  $L_x \times L_z= 6.5 \times 6.5$ cm$^2$. The optical power is $P_S=630$ mW/1 cm$^2$ on the sample, and $P_R=650$ mW/0.32 cm$^2$ on the GaAs crystal. Acoustic pressure is $P_{max}=1.6 MPa$, and lock-in integration time is $\tau_c=100$ ms. Steps along $x$ and $z$ are $\delta_x=0.65$ mm, and $\delta_z=0.65$ mm yielding $100\times 100$ pixels images. The jump time is
$T_{\Phi}=1.9 ~\mu$s  yielding  $\Delta z=$ 2.9 mm. The  data are low pass filtered in the Fourier space. The spatial cutoff frequency is $k_c = $ 1 mm$^{-1}$ for $k_x$ and for $k_z$.

Figure \ref{fig_image_banana} shows the $xz$ image of the sample. Since there is no absorbing inclusion, we get an  image of the  cloud  of photons that travel through the sample and that reach the photodiode. The US beam \\
es the cloud,  and the image corresponds to a  cut of the cloud along the US beam $z$ axis, and  the PZT motion $x$ axis.  As shown on Fig.\ref{fig_image_banana} the cloud is  symmetric by rotation around the $y$ axis,  exhibiting the same extension within the $x$ and $z$ direction as shown on Fig.\ref{fig_cuts_banana}. This is expected since the $x$ and $z$ extension (17.5 and 16.8 mm) is much larger than the resolution along $x$ (FWHM = 1.2 mm for the focused US beam) and $z$ ($\Delta z = 2.9$ mm).

\subsection{Quantitative study of the $z$ resolution in the high resolution case: images and  cuts made with a low scattering sample  ($\mu'_s\simeq 0$) with two absorbing inclusions.}\label{section non diffusing sample}

\begin{figure}[]
\begin{center}
    \includegraphics[width=3 cm]{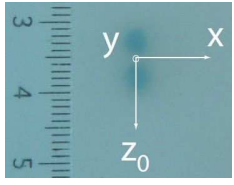}\\
\caption{
Photographs of the non scattering sample
($e= 3.0$ cm, $ \mu'_s =0$)  having 2 absorbing  inclusions (diameter: 3 mm, spacing: 2 mm).
}
\label{fig_photo_echantillon_resolution}
\end{center}
\end{figure}

\begin{figure}[]
\begin{center}
    \includegraphics[width=8 cm]{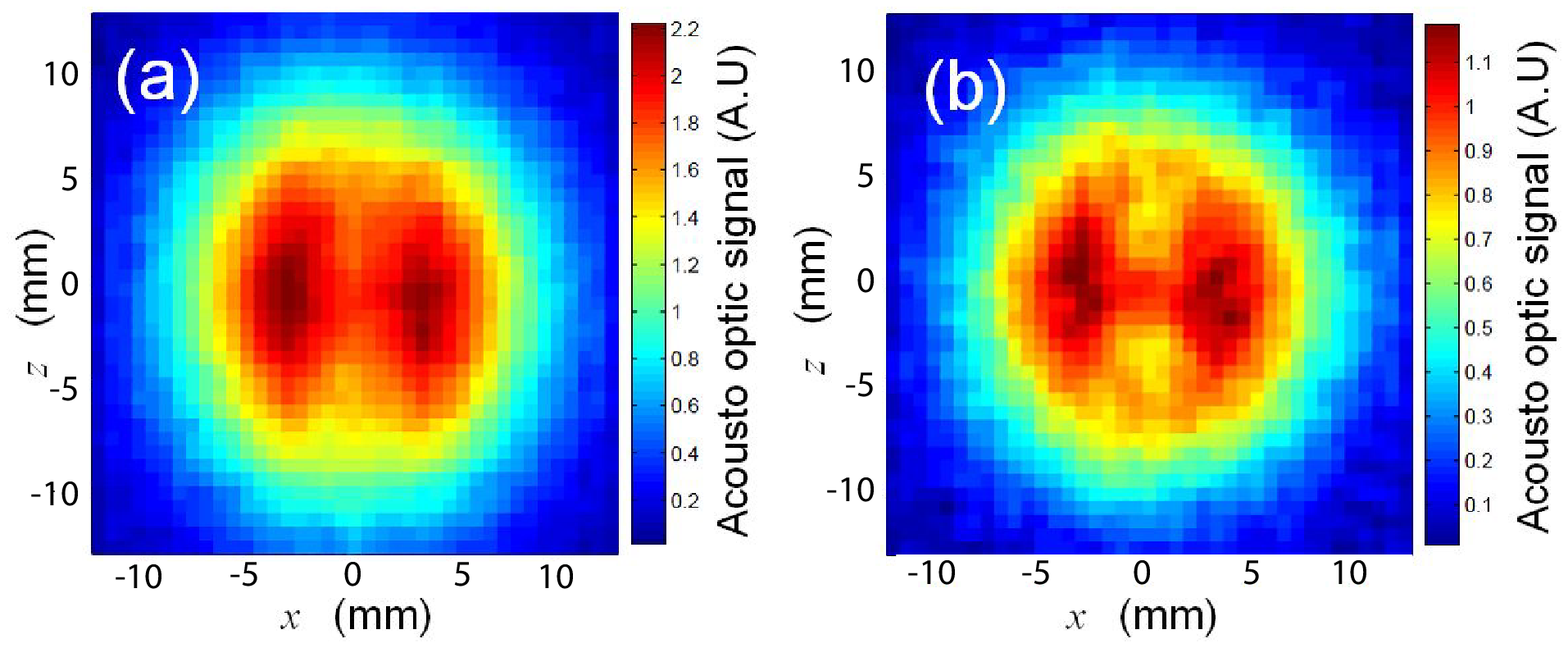}\\
   \includegraphics[width=8 cm]{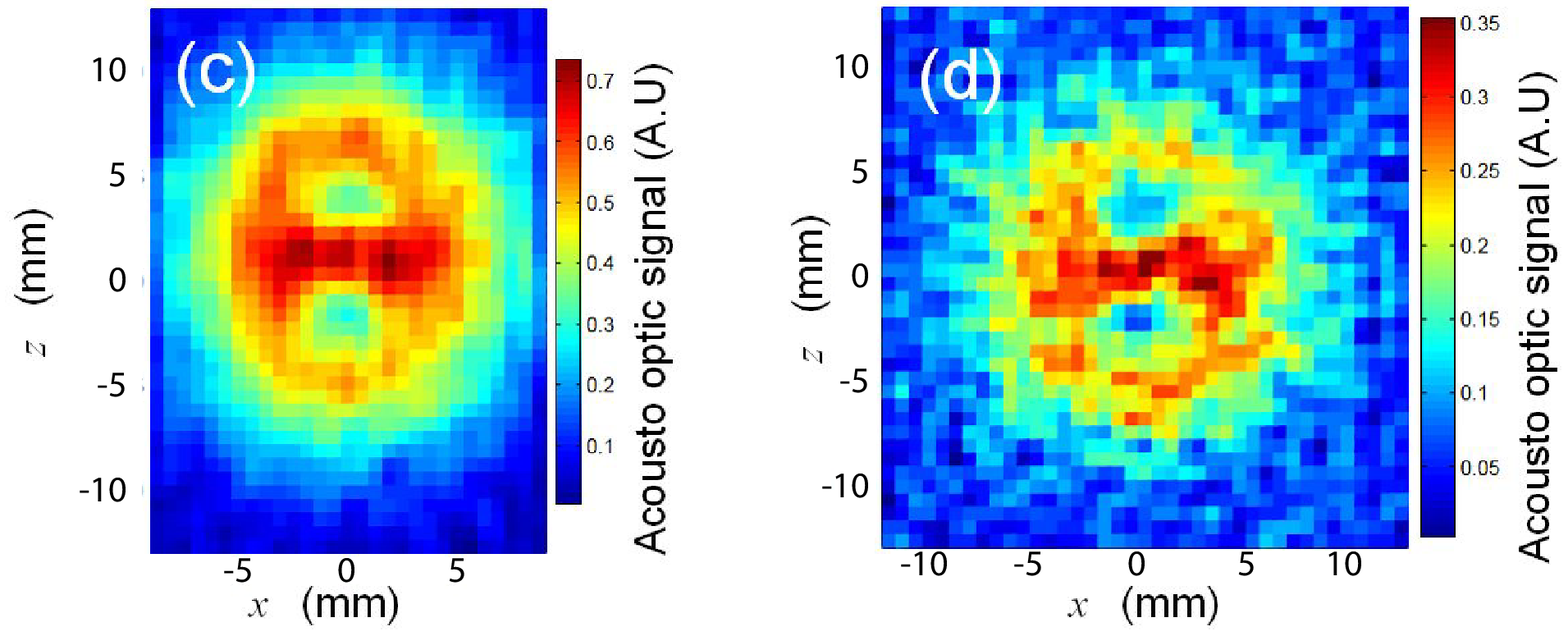}\\
\caption{AOCT $xz$ images  of the sample of Fig.\ref{fig_photo_echantillon_resolution} with $ \Delta z =$  11.4 mm (a), 5.7 mm (b), 2.9 mm (c)  and 1.4mm (d).}
\label{fig_image_echantillon_resolution}
\end{center}
\end{figure}

\begin{figure}[]
\begin{center}
    \includegraphics[width=8.5 cm]{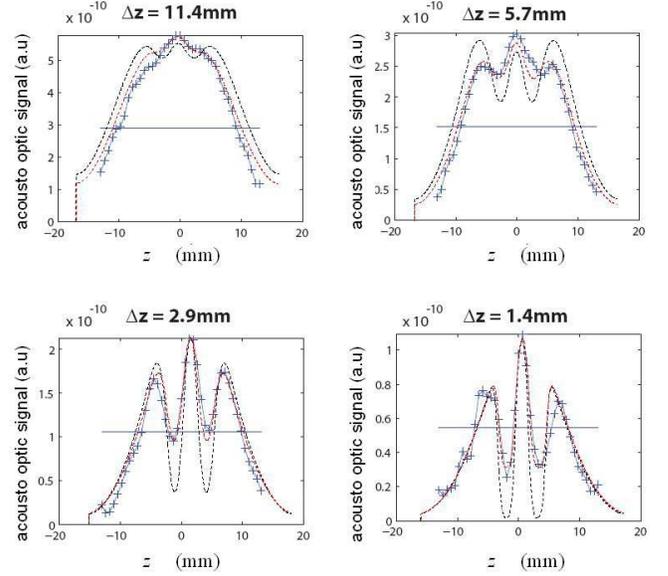}\\
\caption{Profiles of the AOCT photodiode lock-in signal (crosses and blues curves)  of the sample of Fig.\ref{fig_photo_echantillon_resolution} along  $z$ direction with $ \Delta z =$  11.4 mm (a), 5.7 mm (b), 2.9 mm (c)  and 1.4mm (d), and  theoretical profiles (dashed line curves) calculated with totaly (black ) and partially (red) absorbing  inclusions. }
\label{fig_cut_echantillon_resolution}
\end{center}
\end{figure}

To perform a quantitative study of the $z$ resolution in the high resolution case ($\Delta z$ down to 1.4 mm), we considered a sample with two absorbing inclusions of known geometry. These inclusions were made by Agar Agar with the addition of India ink drops. They have a cylindrical shape (diameter: 3 mm, length 5 mm), separated by 2 mm, and  oriented along  $y$ direction (see Fig.\ref{fig_photo_echantillon_resolution}). In pratice, the absorption coefficient was not properly measured but was considered as high. We considered a low scattering sample with Agar Agar and without intralipid. This has been done to get a better sensitivity at high resolution, (since the AOCT signal is proportional to  $\Delta z$) an to we avoid shadow effect. Indeed, the density of photons  near an absorbing inclusion is lower than expected, because the diffused photons that travel through the sample may have crossed or will cross the absorbing zone. The scattering coefficient $\mu'_s$ is nevertheless non zero (as seen on the image of Fig. \ref{fig_photo_echantillon_resolution} ) yielding a photon diffusion cloud, whose  size  is not well known. Since AOCT is based on the detection of diffuse light, a diffuser (sheet of white paper) is placed at the outlet of sample. The experimental conditions are: $P_S=1$ W/ 1 cm$^2$ on the sample; $P_R=630$ mW/0.32 cm$^2$ on the crystal; $P_{max}=1.6$ MPa; $\tau_c=100$ ms; $k_c = $ 1 mm$^{-1}$.

We have studied the $z$ resolution  by imaging the sample and by plotting vertical cuts  for   $\Delta z$ varying from 11.4 mm to 1.4 mm. The images are displayed  on Fig.\ref{fig_image_echantillon_resolution}  with   $\Delta z =$ 11.4 mm (a),  5.7 mm (b), 2.9 mm (c)  and 1.4 mm (d).   Figure \ref{fig_cut_echantillon_resolution} presents their  axial profiles along the vertical axis connecting the two  inclusions (crosses and solid blue curves). On the image corresponding to  $\Delta z =$ 11.4 mm (a),  we cannot  distinguish  any inclusion. We just see  a decrease of the  signal along the vertical column. For  $\Delta z =$  5.7 mm (b),  the inclusions are seen,  but the contrast is low.  Changing to  $\Delta z =$ 2.9 mm (c), the two inclusions are clearly resolved, but the contrast is not yet optimum.
The latter is maximum for  $\Delta z =$  1.4 mm (d), but the  signal becomes quite low, especially  near the absorbing inclusion. Image (d) is thus noisy. Typically, the best compromise between resolution and amplitude is obtained when the resolution is comparable to the dimensions of the object, that is to say for $Delta z = 2.9mm$ compared to the diameter of the inclusion of 3mm. We can also notice from image (a) to (d) that the acousto-optic signal decreases linearly with $\Delta z $.

In order to compare the experimental results to theory, we have also plotted on Figure \ref{fig_cut_echantillon_resolution} the theoretical profiles $S_{PD}(z)$ (dashed curves). The calculation of $S_{PD}(z)=\underline {g_1^2}(z) \otimes
C_z(z)$ is made  by considering that the density of photons $C_z(z)$ within the  inclusions is either zero (black dashed curves) or attenuated by a factor $0.3$ (red dashed curves) with respect to the  density within the cloud. Moreover, the  gaussian profile of the cloud ($\exp(-x^2/w_g^2) $) is supposed to be a little narrower  than in section  \ref{section_z_quant_study} ($w_g=8.0$  in place of 10.2 mm). As seen on Fig. \ref{fig_cut_echantillon_resolution},  the imaging contrast obtained in experiment is lower than the one expected for  totally absorbing inclusions  (black dashed curves). This effect can be related to the imperfect absorption of the inclusion. By considering a finite absorption of 0.3, the agrement of the experiment with theory is much better (red dashed curves).

\subsection{Sample with two absorbing inclusions}

\begin{figure}[]
\begin{center}
    \includegraphics[width=3 cm]{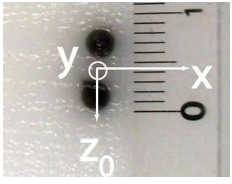}\\
\caption{Pictures  of the diffusing sample  ($e =$ 2 cm, $ \mu'_s = 10 \textrm {cm}^{-1} $) with two absorbing  (3 mm diameter, 2 mm spacing) at mid-thickness  in the plane ($yz$).}
\label{fig_photo_echantillon_diffusant}
\end{center}
\end{figure}

\begin{figure}[]
\begin{center}
    \includegraphics[width=9 cm]{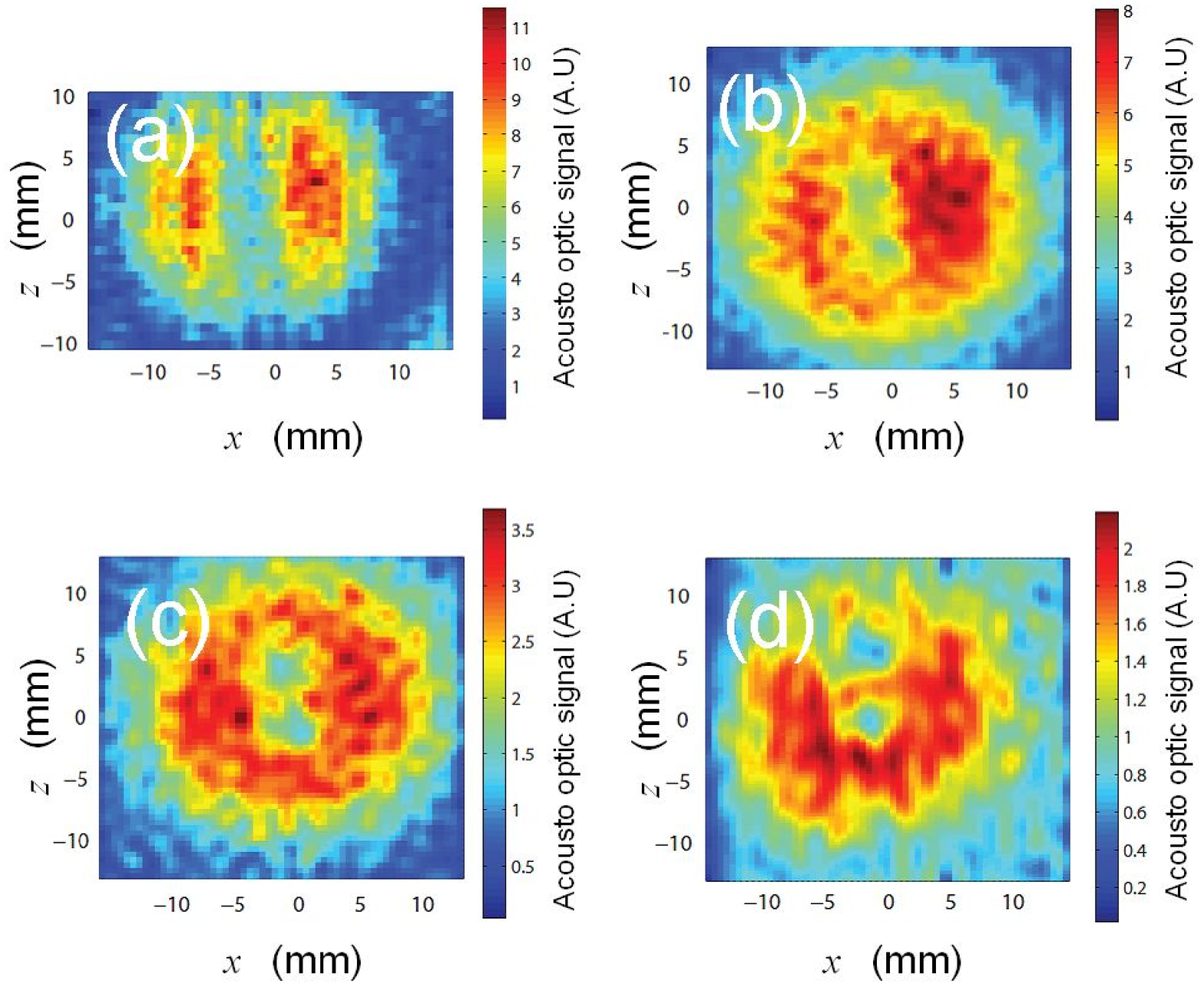}
\caption{AOCT $xz$ images  of the sample of Fig.\ref{fig_photo_echantillon_diffusant} with $\Delta z =$  11.4 mm (a) , 5.7 mm (b) , 2.9 mm (c)  and 1.4 mm (d).
}
\label{fig_image_echantillon_diffusant}
\end{center}
\end{figure}

\begin{figure}[]
\begin{center}
    \includegraphics[width=8.5 cm]{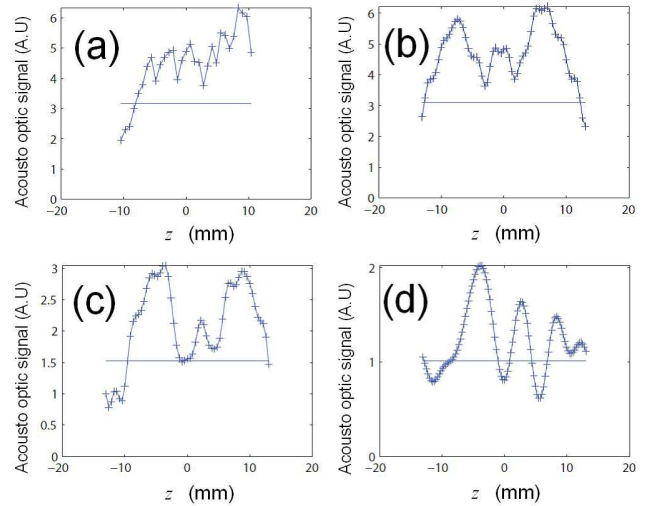}\\
\caption{
Profiles of the acousto optic AOCT photodiode lock-in signal  of the sample of Fig.\ref{fig_photo_echantillon_diffusant} along  $z$ direction with $ \Delta z $ of 11.4 mm (a), 5.7 mm (b), 2.9 mm (c) and 1.4 mm (d).
}
\label{fig_cut_echantillon_diffusant}
\end{center}
\end{figure}

To illustrate  how $\Delta z$ affects the image quality on a more realistic situation, we have imaged a scattering sample with two absorbing inclusions, whose are similar to those used in section \ref{section non diffusing sample}.  These inclusions were made by the same mix of Agar Agar+Intralipides than the rest of the sample, with the addition of India ink drops. They have a cylindrical shape (diameter: 3 mm, length 5 mm), separated by 2 mm, and  oriented along  $y$ direction (see Fig.\ref{fig_photo_echantillon_resolution}). The sample thickness is $L_y=2$ cm, with a scattering coefficient  $\mu'_s= 10 \textrm{cm}^{-1}$. Two inclusions of 3mm diameter are oriented along the $y$ direction  and are spaced 2 mm (see Fig.\ref{fig_photo_echantillon_diffusant}). The experimental conditions are: $P_S=350$ mW/ 1 cm$^2$ on the sample; $P_R=650$ mW/0.32 cm$^2$ on the crystal; $P_{max}=1.6$ Mpa; $\tau_c=100$ ms; $k_c = $ 1 mm$^{-1}$.

Figure \ref{fig_image_echantillon_diffusant}   shows the experimental images obtained for $\Delta z =  11.4$ mm, 5.7 mm,  2.9 mm and 1.4 mm.  Figure \ref{fig_cut_echantillon_diffusant}  shows the corresponding  vertical cut along $z$, at the center of the sample. For $\Delta z =$ 11.4 mm, we do not discern any inclusion, but only a decrease in the signal at the center of the sample that informs us that it is probably one or more absorbing inclusion. For $\Delta z=$ 5.7mm, both inclusions are discernible. This is accentuated for $\Delta z=$ 2.9 mm and 1.4 mm. We can briefly compare these images in scattering medium to those performed in the non-diffusing in previous section \ref{section non diffusing sample}.

For $\Delta z$ = 2.9 mm, the inclusions are less visible here with  scattering  ($\mu'_s= 10 \textrm{cm}^{-1}$),  than without scattering ($\mu'_s= 0$). This can be interpreted as a shadow effect that can occur in  scattering medium. Here again, we can also notice that the signal decreases linearly with $\Delta z$.

As in the section \ref{section non diffusing sample}, the best compromise between resolution and amplitude is obtained when the resolution is comparable to the dimensions of the object, that is to say for $Delta z = 2.9mm$ compared to the diameter of the inclusion of 3mm.

\section{Conclusion}

In this paper, we have confirmed that the  Lesaffre et
al. \cite{lesaffre2009acousto} AOCT technique is able to bring $z$ resolution to acousto optic imaging. We have presented new experimental results, and we have studied  how the $z$ resolution depends on the jumping time $T_\phi$, or on the correlation  length  $\Delta z$. By working at low resolution,  we  have performed a quantitative comparison of the measured  $z$ shape with the  theoretical one, and we have shown  by this way that the  theoretical expression of the AOCT Point Spread Function of reference \cite{lesaffre2011theoretical} agrees with experiments.

We have imaged   samples with 2  absorbing inclusion (3 mm diameter, 2
mm spacing) embedded within 2 different samples (diffusing
or not).  The non diffusing sample has been used to analyse the resolution and to compare it quantitatively with theory in the high resolution regime ($\Delta z = 1.4 and 2.9 mm$). The experimental resolution agrees with theory. The scattering sample has been considered to get qualitative images closer to what expected in living sample.  The resolution varies similarly with  $\Delta z$ but is slightly lower. We have interpreted this decreased resolution as a shadow effect.

The non-scattering sample permits to validate the model by avoiding the problem of shadow effect which will tend to decrease the resolution. The corresponding images permit a quantitative analysis of the resolution in the case of a weak scattering sample with a good signal to noise ratio. Otherwise, the study with a high scattering sample show more realistic results, but in return more qualitative because the resolution is limited by the signal to noise ratio and the shadow effect.

Note that the experiments have been done here with \emph{Agar + Intralipid} phantoms, and not with ex-vivo tumor tissues because phantoms are easier to prepare, and because their optical properties (absorption and  reduced scattering coefficients $\mu'_s$)  are more reproducible, and can be thus more precisely measured.  The experimental analysis done here correspond to AOCT experiments made with photorefractive detection of the acousto optic signal  \cite{lesaffre2009acousto}, which have been theoretically analyzed  by Lesaffre et al. \cite{lesaffre2011theoretical}. We did not try to analyse in detail the recent AOCT experiments made with holographic detection of the acousto optic signal \cite{benoit2012acousto} because the corresponding theory has not been published yet. Moreover, the comparison of the  figure of merit of the two detection schemes (photorefractive versus holographic) in a clinical in vivo application has not been done.


In order to use the AOCT technique  for biomedical applications, as for example to  get information on  breast tumors, it is necessary to work with thicker scattering samples ($>5$ cm), while respecting the acoustic biomedical standards.
In practice, this is very difficult to realize at  $\lambda = 1064$ nm because of water absorption.
On the other hand, experiments realized within the optical therapeutic window - i.e. around $\lambda = 800$ nm -  are easier because the absorption is lower and therefore the transmitted signal much bigger.
This change of optical wavelength involves the use of other photorefractive  crystals (as SPS:Te, ZnTe, BSO) whose development is in progress \cite{farahi2010photorefractive,salma2012trta}.
Besides, the biomedical safety recommendations concern both the optical and acoustical powers \cite{biomedical_standard_iec,biomedical_standard_FDA}. For the optical power, the limit is 1W per cm² at $\lambda=1064nm$ for optical continuous emission, and all the experiments in this paper follow the standard. The acoustic power standards concern the peak to peak acoustic power and the average acoustic power. The first one requires a maximum instantaneous power of $190W/cm^2$, corresponding to a maximum pressure of $2.5MPa$ in water. All the experiments in this paper follow this first acoustic standard. The second acoustic standard requires an average acoustic power of $720mW/cm^2$. Because of the acoustic continuous emission of this set-up, this second acoustic is not reached.
To satisfy this last standard, one could use the AOCT technique with long acoustic pulses (about 1 ms), which are compatible with the long pulse lasers already used in Ultrasound-modulated optical imaging \cite{rousseau2008umo}.
These two modifications would allow the AOCT technique  to be used with thicker samples, while respecting the acoustic biomedical standards.
%
%

This work was supported by funds from the French National Research Agency (ANR-2011-BS04-017-SIMI
4-ICLM).

\bibliographystyle{unsrt}



\end{document}